\documentclass[aps,twocolumn,prb,longbibliography,showpacs,floatfix,superscriptaddress]{revtex4-2}

\usepackage{graphicx,color}
\usepackage{amsfonts}
\usepackage[figuresright]{rotating}
\usepackage{amssymb}
\usepackage{bm}
\usepackage{amsmath}
\usepackage{mathtools}
\usepackage[colorinlistoftodos,prependcaption,textsize=tiny]{todonotes}
\usepackage{xargs}
\usepackage{psfrag}
\usepackage{floatrow}
\usepackage{multirow}
\usepackage{tabularx}
\usepackage{textcomp}
\usepackage{units}
\usepackage{lipsum}
\usepackage{soul}
\usepackage{titlesec}
\usepackage{times}
\usepackage{hyperref}
\usepackage{enumitem}
\usepackage{caption}
\usepackage{subcaption}
\usepackage{xcolor}
\DeclareMathAlphabet\mathbfcal{OMS}{cmsy}{b}{n}
\graphicspath{Images}
\titlespacing*{\section}
{0pt}{1ex plus .25ex}{1ex plus 1ex}
\titlespacing*{\subsection}
{0pt}{1ex plus .25ex}{1ex plus 1ex}
\titlespacing*{\subsubsection}
{0pt}{1ex plus .25ex}{1ex plus 1ex}

\titleformat{\section}
 {\fontsize{10}{15}\bfseries }
  {\thesection}
  {1em}
  {}

\titleformat{\subsection}
  {\bfseries }
  {\thesubsection}
  {2em}
  {}

\titleformat{\subsubsection}
  {\itshape}
  {\thesubsubsection}
  {2em}
  {}

\def\beq{\begin{eqnarray}}
\def\eeq{\end{eqnarray}}

\let\baraccent=\= % rename builtin command \= to \baraccent
\renewcommand{\=}[1]{\stackrel{#1}{=}} % for putting numbers above =
\newcommand{\bk}{\boldsymbol{k}} % bold k
\newcommand{\mc}[1]{\mathcal{ #1}} % mathcal
\makeatletter

\titleclass{\subsubsubsection}{straight}[\subsection]

\newcounter{subsubsubsection}[subsubsection]
\renewcommand\thesubsubsubsection{\thesubsubsection.\arabic{subsubsubsection}}
\titleformat{\subsubsubsection}
   {\normalfont\normalsize\itshape \centering}{\thesubsubsubsection}{1em}{}
\titlespacing*{\subsubsubsection}
{0pt}{1ex plus .3ex}{1ex plus .3ex}
\makeatletter
\renewcommand\paragraph{\@startsection{paragraph}{5}{\z@}%
  {3.25ex \@plus1ex \@minus.2ex}%
  {-1em}%
  {\normalfont\normalsize}}
\renewcommand\subparagraph{\@startsection{subparagraph}{6}{\parindent}%
  {3.25ex \@plus1ex \@minus .2ex}%
  {-1em}%
  {\normalfont\normalsize}}
\def\toclevel@subsubsubsection{4}
\def\toclevel@paragraph{5}
\def\toclevel@paragraph{6}
\def\l@subsubsubsection{\@dottedtocline{4}{7em}{4em}}
\def\l@paragraph{\@dottedtocline{5}{10em}{5em}}
\def\l@subparagraph{\@dottedtocline{6}{14em}{6em}}
\makeatother

\newcommandx{\unsure}[2][1=]{\todo[linecolor=red,backgroundcolor=red!25,bordercolor=red,#1]{#2}}
\newcommandx{\change}[2][1=]{\todo[linecolor=blue,backgroundcolor=blue!25,bordercolor=blue,#1]{#2}}
\newcommandx{\info}[2][1=]{\todo[linecolor=green,backgroundcolor=green!25,bordercolor=green,#1]{#2}}
\newcommandx{\improvement}[2][1=]{\todo[linecolor=yellow,backgroundcolor=yellow!25,bordercolor=yellow,#1]{#2}}
\newcommandx{\thiswillnotshow}[2][1=]{\todo[disable,#1]{#2}}

\begin{document}
\title{Multiplicative Chern Insulator}
\author{Archi Banerjee}
\affiliation{Max Planck Institute for Chemical Physics of Solids, Nöthnitzer Strasse 40, 01187 Dresden, Germany}
\affiliation{Max Planck Institute for the Physics of Complex Systems, Nöthnitzer Strasse 38, 01187 Dresden, Germany}
\affiliation{SUPA, School of Physics and Astronomy, University of St Andrews, St Andrews KY16 9SS, United Kingdom}

\author{Ashley M. Cook}
\affiliation{Max Planck Institute for Chemical Physics of Solids, Nöthnitzer Strasse 40, 01187 Dresden, Germany}
\affiliation{Max Planck Institute for the Physics of Complex Systems, Nöthnitzer Strasse 38, 01187 Dresden, Germany}
\begin{abstract}
    We study multiplicative Chern insulators (MCIs) as canonical examples of multiplicative topological phases of matter. Constructing the MCI Bloch Hamiltonian as a symmetry-protected tensor product of two topologically non-trivial parent Chern insulators (CIs), we study two-dimensional (2D) MCIs and introduce 3D mixed MCIs, constructed by requiring the two 2D parent Hamiltonians share only one momentum component. We study the 2D MCI response to time reversal symmetric flux insertion, observing a $4\pi$ Aharonov-Bohm effect, relating these topological states to fractional quantum Hall states via the effective field theory of the quantum skyrmion Hall effect. As part of this response, we observe evidence of quantisation of a proposed topological invariant for compactified many-body states, to a rational number, suggesting higher-dimensional topology may also be relevant. Finally, we study effects of bulk perturbations breaking the symmetry-protected tensor product structure of the child Hamiltonian, finding the MCI evolves adiabatically into a topological skyrmion phase.
\end{abstract}
\maketitle

%As well, numerics indicate an additional topological invariant associated with compactified many-body states---and intrinsically higher-dimensional topology---also plays a role in this response.

The quantum Hall effect (QHE)~\cite{PhysRevLett.45.494, PhysRevLett.48.1559} revealed the powerful role of topology in modern condensed matter physics~\cite{laughlin1981,halperin1982quantized, Haldane:1983xm, niu1985, PhysRevLett.50.1395, PhysRevLett.62.82, PhysRevLett.63.199, PhysRevLett.64.1313}. Its lattice counterpart, the Chern insulator (CI)~\cite{haldane1988model}, is the prototypical topological insulator~\cite{kane2005, fu_topological_2007, moore2007, QSHI-HgTe-Theory}, and still the only one with quantisation precision of transport signatures comparable to that of the QHE itself~\cite{chang2013, koenig2007, hsieh_topological_2008, chen2009, chen2010, peng_aharonovbohm_2010, liu2014, liu_stable_2014, xu2015, tang_quantum_2017, nadjperge2014, roushan_topological_2009, xia_observation_2009, huang_weyl_2015}. The CI is the foundation of rich classification of band topology~\cite{Ryu_2010, schnyder2008, kitaev_periodic_2009}, but also the foundation for many strongly-correlated states~\cite{ANDERSON1973153, ANDERSON87, BASKARAN1987973, PhysRevLett.59.2095, PhysRevB.35.8865, PhysRevLett.61.2376, PhysRevB.37.3774, PhysRevLett.66.1773, PhysRevB.40.7387, PhysRevB.44.2664, PhysRevB.62.7850, PhysRevLett.86.292, wegner1971duality, RevModPhys.51.659, PhysRevLett.86.1881, KITAEV20032, KITAEV20062}.

The CI therefore represents a dichotomy between non-interacting and correlated states, motivating the present study of multiplicative topological phases (MTPs)~\cite{cook_multiplicative_2022} within the framework of the quantum skyrmion Hall effect (QSkHE)~\cite{cook2023, cook_multiplicative_2022, cookFST2023, qskhe, patil2024}, as these phases are characterized by symmetry-protected tensor product Hilbert spaces similar to the Fock space of multiple particles.  We study the MCI as a canonical example~\cite{cook_multiplicative_2022}. In addition to introducing higher-dimensional MCIs, we study MCI response signatures and robustness of the MCI against symmetry-breaking perturbations. We observe a $4 \pi$ Aharonov-Bohm (AB) effect consistent with Chern-Simons (CS) theories similar to those of the fractional quantum Hall effect (FQHE)~\cite{PhysRevLett.50.1395, PhysRevLett.64.1313, PhysRevLett.63.199, kalmeyer1989, halperin1993, suen1992, PhysRevLett.62.82}. We also find MCIs can evolve adiabatically into topological skyrmion phases of matter upon breaking the bulk product structure~\cite{cook2023, calderon_skyrm, winterOEPT, liu2020}.

\maketitle
\begin{figure}[tbh!]
  \centering
 % \subfloat[]{\includegraphics[height=0.35\textwidth,width=0.5\textwidth]{Images/Fig1/CICartoon1.png}\label{fig:f11}}
  %\subfloat[]{\includegraphics[width=0.5\textwidth]{Images/Fig1/1b.png}\label{fig:f12}}
  %\vfill
  %\subfloat[]{\includegraphics[width=0.5\textwidth]{Images/Fig1/1c.png}\label{fig:f13}}
  %\subfloat[]{\includegraphics[width=0.5\textwidth]{Images/Fig1/1d.png}\label{fig:f14}}
  \includegraphics[width=\textwidth]{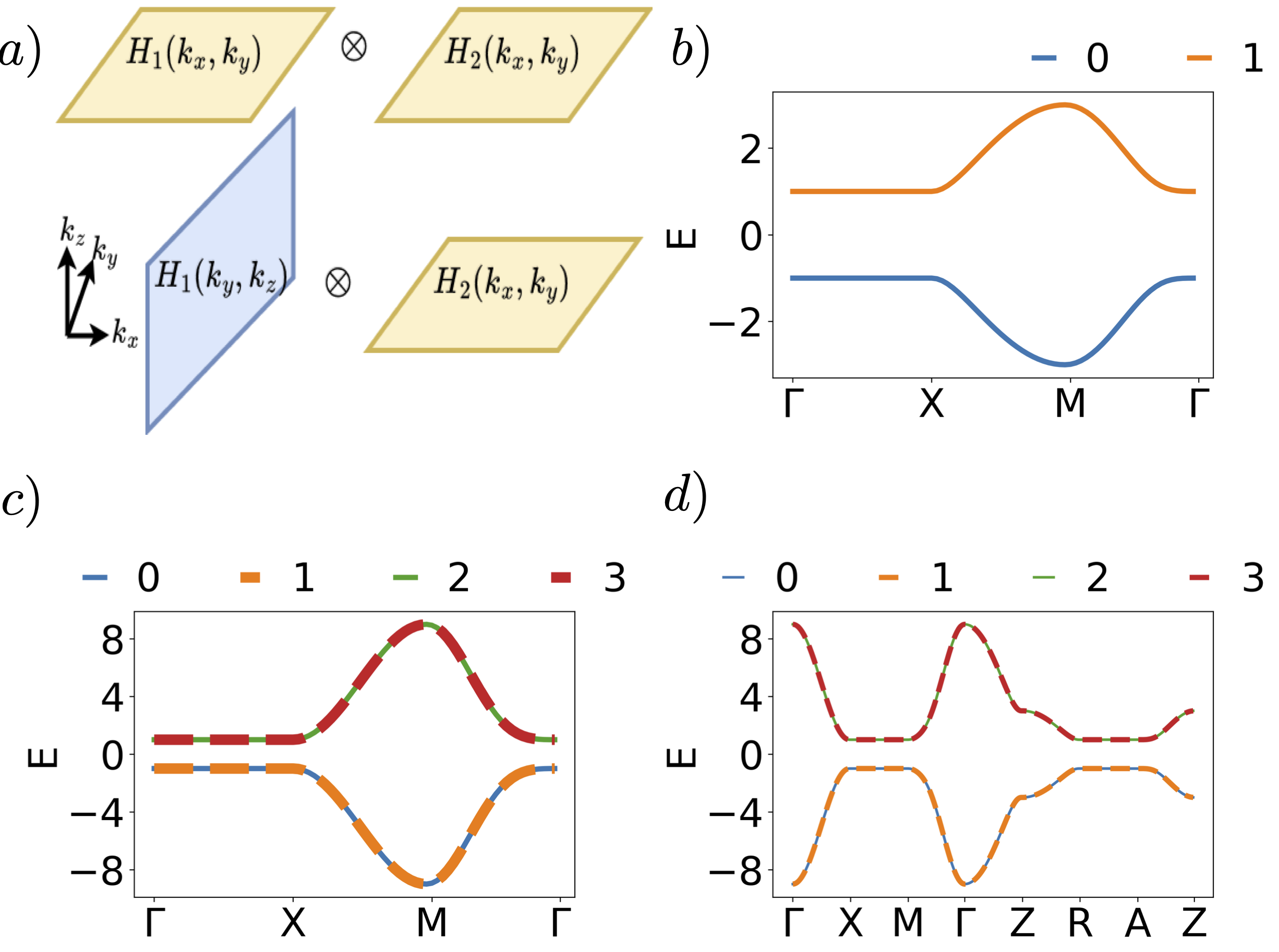}
  \caption{ a) Schematic depicting construction of 2D (3D) MCI as tensor product of two parent Hamiltonians sharing two (one) momentum component, b-d) bulk band structure for b) a parent Hamiltonian with $(m,t,\triangle)=(+1,+0.5,-1)$, c) the 2D MCI, with both parents sharing $(m,t,\triangle)=(+1,+0.5,-1)$, d) the 3D MCI with both parents sharing the parameters $(m,t,\triangle)=(+1,-0.5,+1)$.}
  \label{fig:F1}
\end{figure}

\textit{Model Hamiltonians}---Multiplicative topological phases are realized in Hamiltonians, which are quadratic in creation and annihilation operators. We consider those of the form~\cite{cook_multiplicative_2022}
\begin{align}
    \mathcal{H}_c = \sum_{\boldsymbol{k}_1, \boldsymbol{k}_2} \Psi^{\dagger}_{\boldsymbol{k}_1, \boldsymbol{k}_2}\mathcal{H}^c(\boldsymbol{k}_1, \boldsymbol{k}_2) \Psi^{}_{\boldsymbol{k}_1, \boldsymbol{k}_2},
\end{align}
which are four-band child Bloch Hamiltonians, writing the basis vector as $\Psi_{\boldsymbol{k}_1, \boldsymbol{k}_2} = \left( c^{}_{\boldsymbol{k}_1, \boldsymbol{k}_2, \uparrow}, c^{}_{\boldsymbol{k}_1, \boldsymbol{k}_2, \downarrow}, c^{\dagger}_{-\boldsymbol{k}_1, -\boldsymbol{k}_2, \downarrow}, c^{\dagger}_{-\boldsymbol{k}_1, -\boldsymbol{k}_2, \uparrow}\right)^{\top}$, where $c^{\dagger}_{\boldsymbol{k}_1, \boldsymbol{k}_2, \sigma}$ creates a fermion labeled by two momenta $\boldsymbol{k}_1$ and $\boldsymbol{k}_2$ and spin $\sigma$. The multiplicative child Bloch Hamiltonian is expressed in terms of two parent Bloch Hamiltonians $\mc{H}_{p1}(\boldsymbol{k}_1)$ and $\mc{H}_{p2}(\boldsymbol{k}_2)$ combined by Kronecker product. This product structure is symmetry-protected~\cite{cook_multiplicative_2022}. The parent Hamiltonians $\mc{H}_{p1}(\boldsymbol{k}_1)$ and $\mc{H}_{p2}(\boldsymbol{k}_2)$ are given as,
\begin{equation}
\begin{split}
&\mc{H}_{p1}(\boldsymbol{k}_1)=\mathbf{d}_1 (\boldsymbol{k}_1)\cdot \boldsymbol{\tau};\quad \mc{H}_{p2}(\boldsymbol{k}_2)=\mathbf{d}_2 (\boldsymbol{k}_2) \cdot \boldsymbol{\sigma},
\end{split}
\end{equation}
where $\mathbf{d}_1(\boldsymbol{k}_1)$ and $\mathbf{d}_2(\boldsymbol{k}_2)$ are momentum-dependent, three-component vectors of scalar functions, each of $\boldsymbol{\sigma}$ and $\boldsymbol{\tau}$ is a vector of Pauli matrices, and $\boldsymbol{k}_{1,2}$ are each momentum vectors with two components.  We take the $\boldsymbol{d}$-vector of parent $1$, $\boldsymbol{d}_1 = \langle d_{11}, d_{21}, d_{31} \rangle$, to be that of the QWZ model~\cite{qi2006_QWZmodel}, $d_{11} = \Delta \sin{k_y}$, $d_{21} = \Delta  \sin{k_x}$, $d_{31} = m-2t\left( \cos(k_x) + \cos(k_y) \right)$. We take $\mathcal{H}_{p2}(\bk_2)$ to be $\mathcal{H}_{p1}(\bk_1)$ subjected to time reversal, using the time-reversal operator $\mathcal{T} = i \sigma_y \mathcal{K}$, where $\sigma_y$ is the second Pauli matrix and $\mathcal{K}$ is complex conjugation, as  $\mathcal{H}_{p2}(\bk) = \mathcal{T} \mathcal{H}_{p1}(\bk)  \mathcal{T}^{-1}$. The child Bloch Hamiltonian is then expressed as $\mc{H}^c(\boldsymbol{k})=(d_{11},d_{21},d_{31})\cdot\boldsymbol{\tau}\otimes (-d_{12},d_{21},d_{31})\cdot \boldsymbol{\sigma}$. We will consider both a two-dimensional (2D) \textit{parallel} MCI ($||$MCI), by taking $\boldsymbol{k}_2 = \boldsymbol{k}_1 = \left(k_x,k_y \right)$. Parents $1$ and $2$ do not have to be related by such transformations, however, and we also consider generalisation to a 3D \textit{mixed} MCI, by taking $\boldsymbol{k}_1 = \left( k_x, k_y\right)$ and $\boldsymbol{k}_2 = \left( k_y, k_z\right)$.

The tensor-product structure guarantees that the eigenvalues of the child Hamiltonian,
$E^c_{12}(\boldsymbol{k})$, are products of eigenvalues of $\mc{H}_{p1}(\boldsymbol{k})$ i.e., $E_{p1}(\boldsymbol{k})$, and $\mc{H}_{p2}(\boldsymbol{k})$ i.e., $E_{p2}(\boldsymbol{k})$, respectively, or $E^c_{12}(\boldsymbol{k})= \pm E_{p1}(\boldsymbol{k})E_{p2}(\boldsymbol{k})$, and therefore also at least doubly degenerate as shown in Fig.~\ref{fig:F1} (c) and (d). Indeed, comparison between the bulk spectrum of a parent Hamiltonian $\mathcal{H}_{p1}(\bk)$ along high-symmetry lines in the Brillouin zone (BZ) in Fig.~\ref{fig:F1} b), and the bulk spectrum of the $||$MCI shown in
Fig.~\ref{fig:F1} c) shows bandwidth of bands in the child Hamiltonian is the square of corresponding parent bands for the case we consider.

In Fig.~\ref{fig:F1} d), we observe both high-symmetry lines along which bands of the child Hamiltonian disperse like those of an individual parent (compare the band structure along the XZ line for the child Hamiltonian in Fig.~\ref{fig:F1} d) with the band structure along the MX line for the parent Hamiltonian in Fig.~\ref{fig:F1} b)), as well as lines along which the dispersion is the square of that for an individual parent (compare the $\Gamma$Z line in Fig.~\ref{fig:F1} d) with the MX line in Fig.~\ref{fig:F1} b)).

%Before closing this section, we comment on broader symmetry-protection of multiplicative topological phases, to provide intuition for later characterization of the MCI. Symmetry-protection yields a flag manifold structure for the set of deformations of the projector-onto-occupied states of the child Hamiltonian~\cite{cook_multiplicative_2022}. This corresponds to a second interpretation of the child Hamiltonian as purely quartic in creation and annihilation operators, with each parent being a single-particle system and the quasiparticle species of parent 1 being indistinguishable from that of parent 2 (encoded by the tensor product), although the parent Hamiltonians can differ.

\begin{figure}[htb!]
  \centering
   % \subfloat[]{\includegraphics[width=0.49\textwidth]{Images/Fig2/QWZSpectra.png}}\label{fig:f21}
   % \subfloat[]{\includegraphics[width=0.49\textwidth]{Images/Fig2/MCI2DSpectra.png}\label{fig:f22}}
   %  \vspace{\fill}
   %  \subfloat[]{\includegraphics[width=0.50\textwidth]{Images/Fig2/MCI3DSpectraX.png}\label{fig:f23}}
   %  \subfloat[]{\includegraphics[width=0.50\textwidth]{Images/Fig2/MCI3DSpectraZ.png}\label{fig:f24}}
   \includegraphics[width=\textwidth]{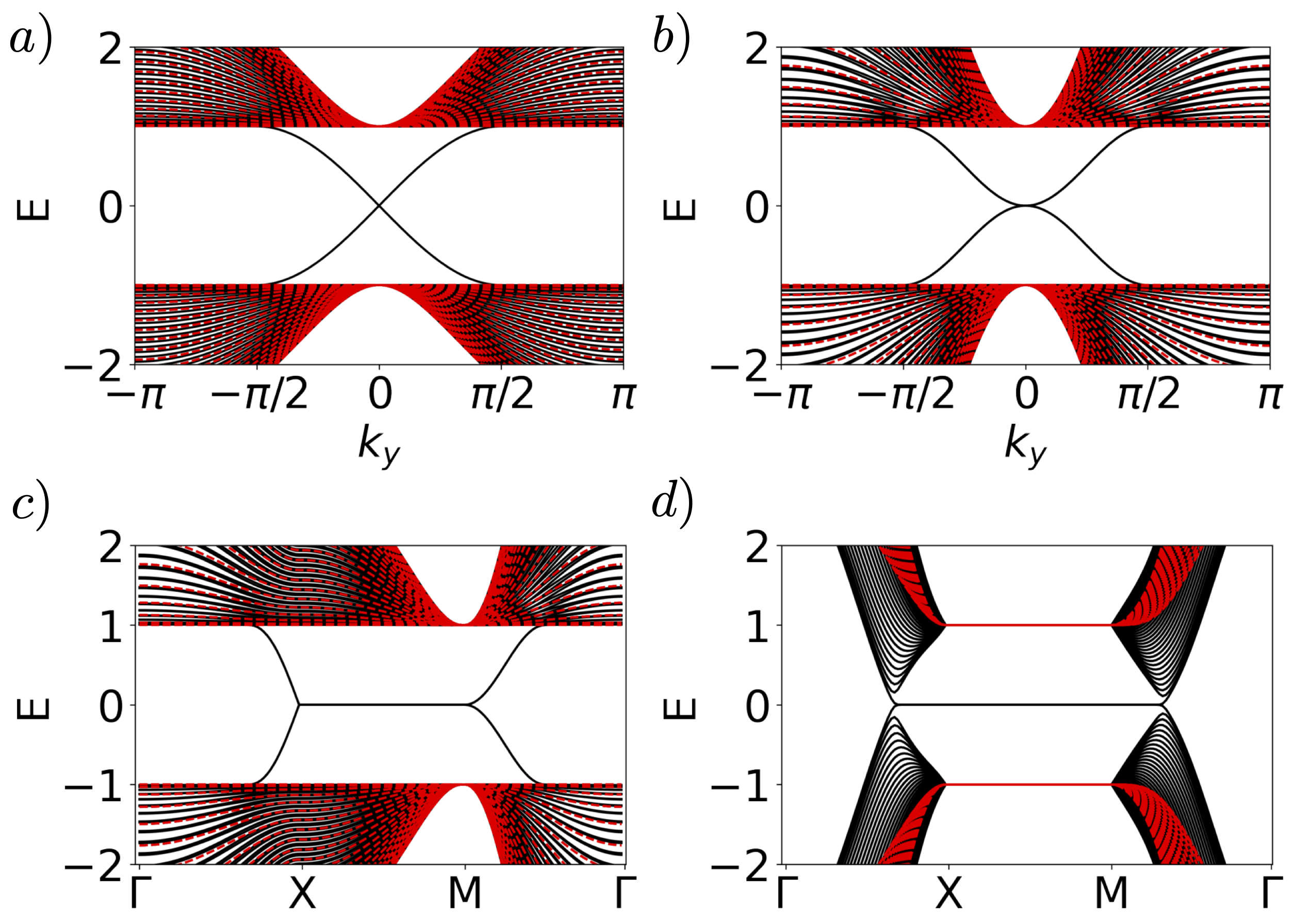}
  \caption{ a) Slab spectrum for a single parent Bloch Hamiltonian with $(m,t,\triangle)=(+1,+0.5,-1)$ for OBCs in the $\hat{x}$-direction, with system size $L_x = 50$. b) slab spectrum for $||$MCI as shown, with both parents having $(m,t,\triangle)=(+1,+0.5,-1)$, c) slab spectrum for mixed MCI for OBCs in $\hat{x}$- (shared) direction, with two parents sharing the parameters $(m,t,\triangle)=(+1,-0.5,+1)$, d) slab spectrum for 3D mixed MCI for OBC in z (unshared direction) with two parents sharing the parameters $(m,t,\triangle)=(+1,-0.5,+1)$}
  \label{fig:F2}
\end{figure}

\textit{Bulk-boundary correspondence}---We now characterize bulk-boundary correspondence of the $||$ and mixed MCIs, with results shown in Fig.~\ref{fig:F2}. The slab spectrum for a single parent is shown in Fig.~\ref{fig:F2} (a) for direct comparison with the slab spectrum for the $||$MCI with open boundary conditions (OBCs) in the $\hat{x}$-direction in Fig.\ref{fig:F2} (b). While the edge modes disperse linearly for an individual parent and are non-degenerate, the edge modes of the $||$MCI disperse quadratically, reflecting a multiplicative dependence of eigenvalues for the child Hamiltonian on parent eigenvalues as in the bulk.  These edge states are exponentially-localized. For $k_y=-\left(4/9\right) \pi$ and system size in the direction of OBCs of $L=50$, we observe boundary modes with probability density peak at layer indices $x=0$, $x=1$, $x=49$, and $x=50$, respectively. Similar edge state probability density peak locations are also observed for the
parallel multiplicative Kitaev chain~\cite{pal2023multiplicative}.

For the 3D mixed MCI, we consider both OBCs in the $\hat{x}$- (shared) direction, given that each of parent 1 and 2 depend on $k_x$, as well as OBCs in the $\hat{z}$- (unshared) direction, given that only one parent depends on $k_z$. Slab spectra for these two cases are shown in Fig.~\ref{fig:F2} (c) and (d), respectively. In Fig.~\ref{fig:F2} (c), the surface bands are everywhere at least two-fold degenerate, and traverse the bulk gap. There is a zero-energy manifold of states along the XM line, which disperses linearly from X towards $\Gamma$ and quadratically from M towards $\Gamma$. These states are exponentially-localized, with probability density peaks located in the first or last layers of the slab. For the case of Fig.\ref{fig:F2} (d), we again observe an extended zero-energy manifold, primarily along high-symmetry line XM. Strikingly, while the system is gapped in the bulk for PBCs, OBCs for an unshared direction yield additional \textit{bulk} states not present for PBCs. The edge states emerge from bulk gap-closings, \textit{which are only present for OBCs}. That is, in this scenario, the system is a topological insulator in the bulk for PBCs, but \textit{a topological semimetal in the bulk for OBCs}.

\begin{figure}[tbh!]
  \centering
  %  \subfloat[]{\includegraphics[height=0.45\textwidth,width=0.50\textwidth]{Images/Fig3/FISchemaFinalForReal.png}\label{fig:f31}}
  % \subfloat[]{\includegraphics[height=0.45\textwidth,width=0.50\textwidth]{Images/Fig3//2DMagResMCIFinal.png}\label{fig:f32}}
  % \vfill
  % \subfloat[]{\includegraphics[height=0.45\textwidth,width=0.50\textwidth]{Images/Fig3/symm0_antisymm1_probdens.png}\label{fig:f33}}
  % \subfloat[]{\includegraphics[height=0.45\textwidth,width=0.55\textwidth]{Images/Fig3/Spin_Charge_plot.png}\label{fig:f34}}
  % \includegraphics[width=\textwidth]{Images/Fig3/Fig3.png}
  %\includegraphics[width=\textwidth]{Images/Fig3/3ab.png}
  %\includegraphics[width=\textwidth]{Images/Fig3/3cd.png}
  \includegraphics[width=\textwidth]{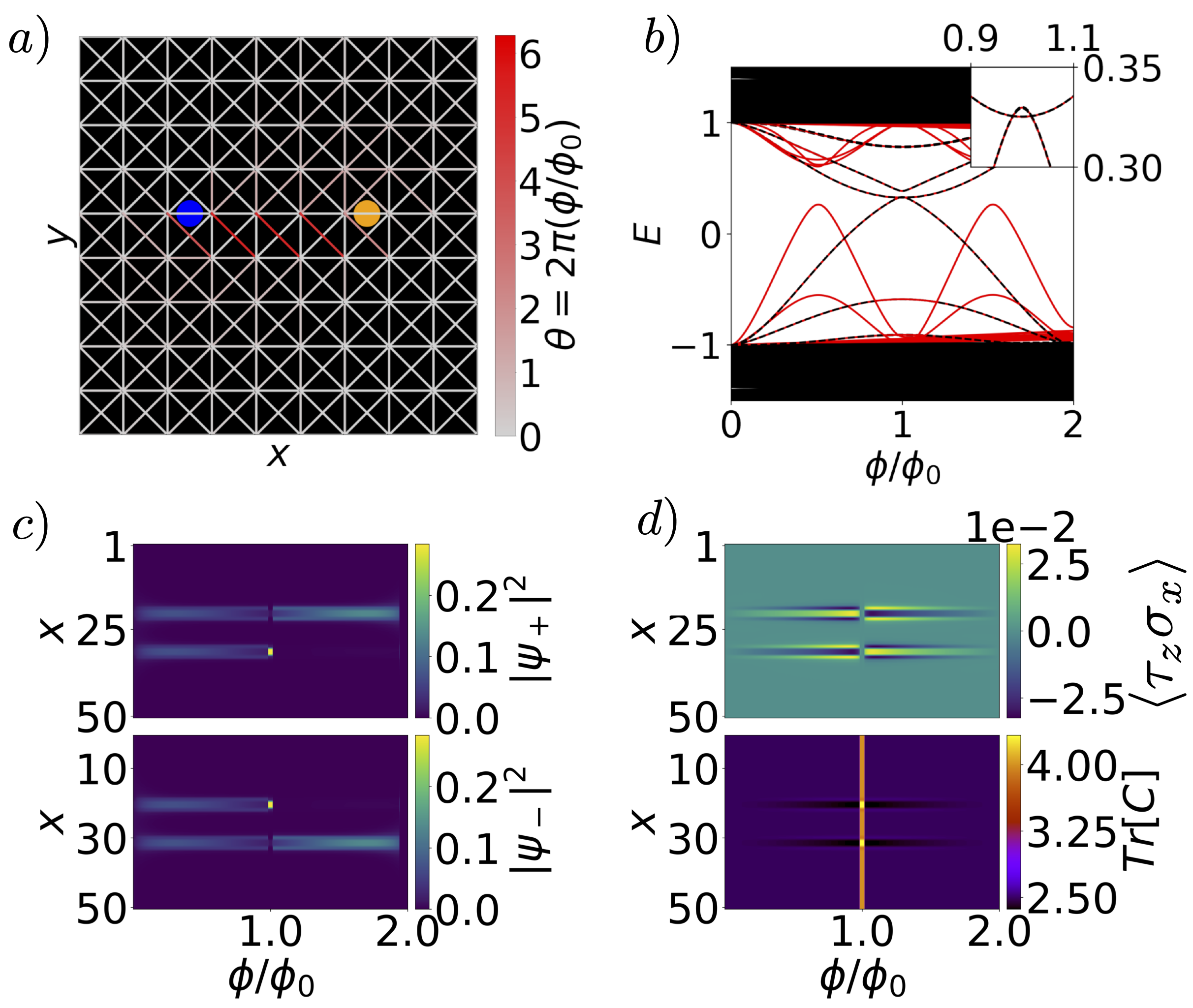}
  \caption{a) TRS flux insertion scheme for a $10 \times 10$ square lattice: TRS pair of flux tubes are located at points marked in orange and blue. The amplitude of Peierl's phases for nearest-neighbor and diagonal hopping integrals are represented by the color bar. b) Energy spectrum of child $||$MCI with $(m,t,\triangle)=(+1,+0.5,-1)$ under flux insertion through a  single plaquette (red) and a pair of plaquettes with opposite fluxes (dashed black) for PBCs, with inset showing level crossings between two-fold degenerate states near $\phi=\phi_0$. c) Evolution upon flux insertion of probability density for states $\psi_{\pm}$ vs. real-space coordinate $x$ ($y_0 = 25$) and magnetic flux $\phi$. d) Ground state expectation value for operator $\tau_z \sigma_x$ (top) and topological charge $Tr[\boldsymbol{C}]$ (bottom).}
  \label{fig:F3}
\end{figure}

\textit{Topological response to external magnetic field}---We now study response signatures of the MCI, focusing on the case of the $||$MCI. In this case, we consider a time-reversal symmetric (TRS) flux insertion at two locations in the lattice for system size $50$ by $50$ unit cells and PBCs. A schematic of this flux insertion for a smaller system size is shown in Fig.~\ref{fig:F3} (a). Evolution of the spectrum vs. magnetic flux $\phi$ is shown in Fig.~\ref{fig:F3} (b), for both TRS flux insertion at two sites of the lattice (black dashed lines), as well as for TRS-breaking flux insertion at just one of these two sites (red). We observe features in the plot of spectrum vs. magnetic flux for flux insertion through a single plaquette, which are $\phi_0$-periodic, where $\phi_0$ is one flux quantum. These correspond to response signatures of the individual parent Hamiltonians, with Chern numbers $C=\pm1$.  Entanglement between the parent degrees of freedom (DOFs) yields gaps near $\phi = \phi_0/2$ and $\phi = 3\phi_0/2$. The spectrum therefore remains gapped until closer to $\phi = \phi_0$, at which point the bulk gap closes in the vicinity of $\phi = \phi_0$ as part of a $2 \phi_0$-periodic feature in the topological response.

Taking the two degenerate eigenstates that enter the bulk gap from the bulk valence bands, outputted by numerical diagonalisation, to be $\phi_{1,2}$, we define particular linear combinations of these states as $\psi_{\pm} = {1 \over \sqrt{2}}\left(\phi_1 \pm \phi_2\right)$. We show evolution of the probability density for $\psi_{\pm}$ for $y_0=25$ vs. $x$ and $\phi$ in Fig.~\ref{fig:F3} c). We observe even distribution of $\psi_{\pm}$ between the sites of flux insertion for $\phi<\phi_0$, and probability density becomes strongly-localised at just one site of flux insertion for each state for $\phi>\phi_0$.

We further characterise evolution of the system under flux insertion by computing the expectation value of $\tau_z \sigma_x$ vs. flux $\phi$ and position $x$ for $y_0=25$, shown in Fig.~\ref{fig:F3} d) (top), observing a net change in sign. $\tau_x \sigma_z$ yields the same result up to a sign reversal. As well, we compute a topological invariant associated with the parent isospin DOFs~\cite{patil2024}. To do so, we construct a spin representation associated with the $\tau$ ($\sigma$) DOF for the occupied states by computing the density matrix from the one-point correlator~\cite{IngoPeschel_2003} for a given site in the lattice as $\rho(x,y)$. We then compute a spin representation for the occupied states as $S^{occ}_{i}(x,y) = \rho(x,y) \tau_i \sigma_0$ ($\tau_0 \sigma_i$ yields the same result). We finally compute the topological charge associated with a compactified many body state as $Tr \left[\boldsymbol{C}(x,y)\right]=Tr \left[\left[ S^{occ}_{x}(x,y) , S^{occ}_{y}(x,y) \right] \left(S^{occ}_{z}(x,y)\right)^{-1}\right]$~\cite{patil2024}.
 Fig.~\ref{fig:F3} (d) (bottom) shows $Tr \left[\boldsymbol{C} \right]$ vs. magnetic flux $\phi$ and layer index in the $\hat{x}$-direction, $x$, for $y_0=25$. At sites of flux insertion for $\phi$ near $\phi_0$, $Tr \left[\boldsymbol{C} \right]$ deviates from $4$, the value expected for the spin representation without projection to the occupied subspace, by very close to $1/3$. This value converges towards $1/3$ with increasing system size and occurs for an extended region of phase space~\cite{SuppMat}.

The $4\pi$ AB effect corresponds to a minimal CS field theory within the framework of the QSkHE~\cite{qskhe, patil2024}. It possesses two types of gauge fields, those coupling to the quasiparticle species of each individual parent DOF, written as U(1) gauge field  $A_{1}$ or U(1) gauge field $A_{2}$, respectively, as well as a gauge field coupling to the total charge of a quasiparticle pair encoded in the isospin DOFs, $a$, with U(1)$\times$U(1) structure. This quasiparticle pair is a quantum skyrmion, which can be modeled as a severely-fuzzified Landau level (LL\textsubscript{F})~\cite{qskhe, patil2024}. This corresponds to a minimal effective action with two terms:
\begin{align}
S_{c,\mathrm{eff}} & = {\frac{C_1}{2 \pi}} \int  d^2x  dt
 \varepsilon^{\mu \nu \rho} A_{1/2,\mu} \partial_{\nu} a_{\rho}\\ \nonumber
 &+{\frac{Q}{4 \pi}} \int  d^2x  dt
 \varepsilon^{\mu \nu \rho} a_{\mu} \partial_{\nu} a_{\rho}.
\end{align}
Here, $A_{1/2,\mu}$ acts on the composite particles by the $A_{1(2),\mu}$ component acting on the parent $1$ ($2$) particle species. The gauge field $\boldsymbol{a}$ may be expressed in terms of the parent gauge fields as $a_{\mu} = A_{1,\mu} + T\left[A_{2,\mu}\right]$, where  $T\left[A_{2,\mu}\right]$ is a chiral transformation on $A_{2,\mu}$. With this definition of the emergent gauge field, $\mathcal{Q}=2$ if $C_1=1$. Here, $\mathcal{Q}$ is the skyrmion number previously-used to characterise topological skyrmion phases of matter~\cite{cook2023, qskhe}. For 2+1 D systems, it takes the form
\begin{equation}
    \mc{Q} = {\frac{1}{4\pi}} \int_{BZ}d\boldsymbol{k} \left[ \langle \boldsymbol{\hat{S}}(\boldsymbol{k}) \rangle \cdot \left(\partial_{k_x}\langle \boldsymbol{\hat{S}}(\boldsymbol{k}) \rangle \times \partial_{k_y}\langle \boldsymbol{\hat{S}}(\boldsymbol{k}) \rangle \right) \right],
    \label{skyrmnum}
\end{equation}
where $\boldsymbol{S} = \left(S_1, S_2, S_3\right)$ is a (pseudo)spin representation specific to the DOFs of a given system,
and $\langle \boldsymbol{\hat{S}}(\boldsymbol{k}) \rangle$ is the normalized
expectation value of the spin for occupied states. Here $\langle
S_i(\boldsymbol{k}) \rangle= \sum_{n \in \mathrm{occ}} \langle n,
\boldsymbol{k} | S_i | n, \boldsymbol{k} \rangle $, with $i \in \{1,2,3\}$ and
$| n, \boldsymbol{k} \rangle$ the Bloch state associated with the
$n$\textsuperscript{th} band.
We can compute $Q$ analytically due to the tensor product structure of the bulk Bloch Hamiltonian as well, finding $Q=2$ for parent Chern numbers of $\pm 1$~\cite{SuppMat}. Then the CS theory has the minimal terms for the $\nu = 1/2$ FQH plateau~\cite{PhysRevLett.62.82} and $4 \pi$ AB effect. However, results for $Tr\left[\boldsymbol{C} \right]$ in Fig.~\ref{fig:F3} (d) suggest an additional topological invariant for compactified many-body states~\cite{patil2024}---and potentially higher-dimensional terms in the CS theory---are needed for deeper understanding of the MCI.

\begin{figure}[tbh!]
  \centering
   % \subfloat[]{\includegraphics[height=0.4\textwidth,width=0.50\textwidth]{Images/Fig4/4a.png}\label{fig:f41}}
 % \subfloat[]{\includegraphics[height=0.4\textwidth,width=0.55\textwidth]{Images/Fig4/4b.png}\label{fig:f42}}
 % \vfill
 % \subfloat[]{\includegraphics[height=0.4\textwidth,width=0.45\textwidth]{Images/Fig4/4c.png}\label{fig:f43}}
  %\subfloat[]{\includegraphics[height=0.4\textwidth,width=0.50\textwidth]{Images/Fig4/4d.png}\label{fig:f44}}
  \includegraphics[width=\textwidth]{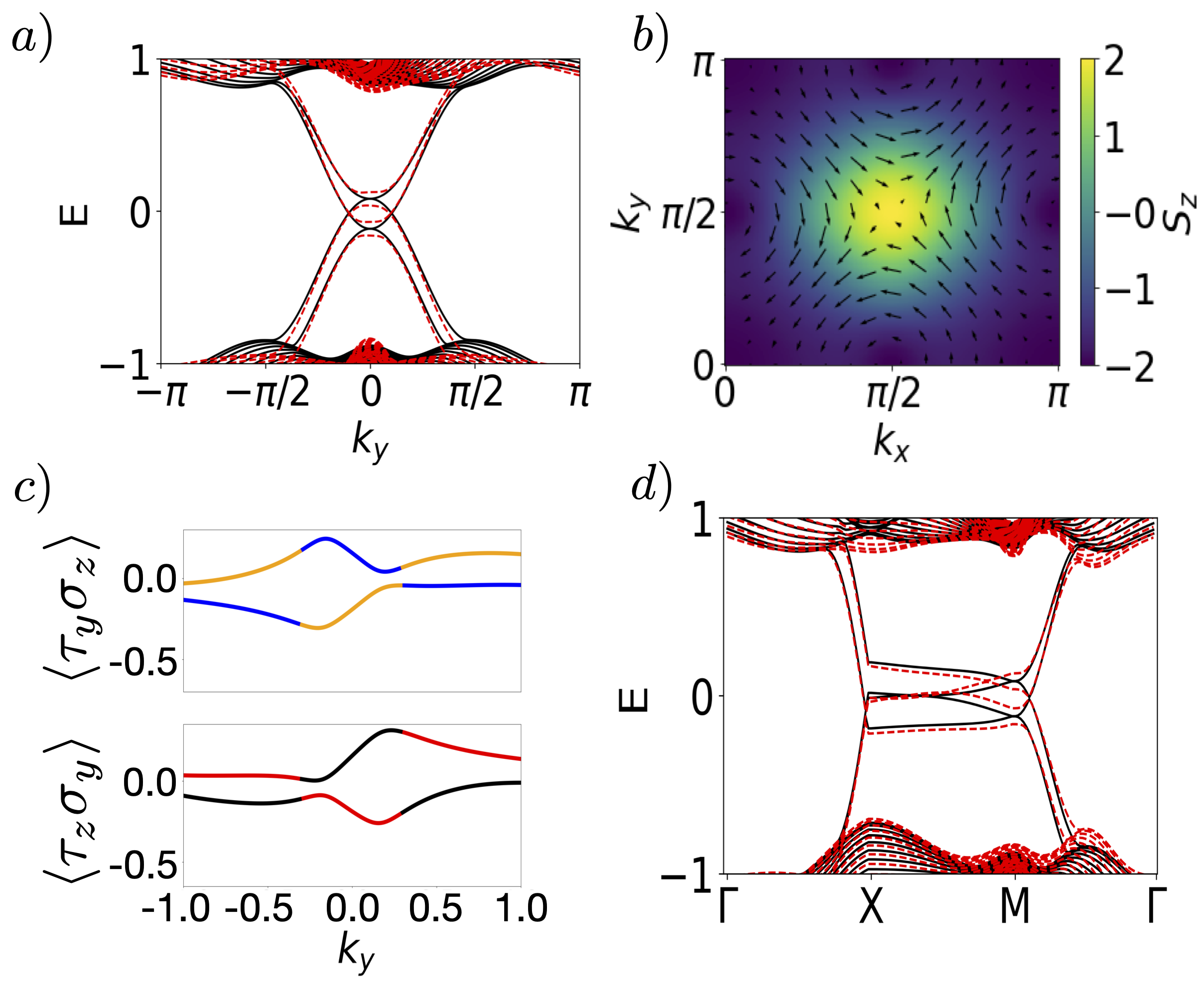}
  \caption{ a) Slab spectrum of $||$MCI (with both parents sharing $(m,t,\triangle)=(+1,+0.5,-1)$), with $m' = 0.1$ and $t'=0$ (black) and $m'=0.1$ and $t'=0.05$ (red), with OBCs (PBCs) in the $\hat{x}$- ($\hat{y}$-) direction and system size in the $\hat{x}$-direction of $L=50$. b) BZ spin texture with skyrmion number $1$ for spin representation $\tau_i \sigma_0$. c)  $\langle\tau_y \sigma_z\rangle$ and $\langle\tau_z \sigma_y\rangle$ for the 99\textsuperscript{th} (blue and red) and 100\textsuperscript{th} (orange and black) lowest-energy eigenstates of the $||$MCI for $m'=0.1$ and $t'=0.05$ vs. $k_y$, d) Slab spectrum of 3D mixed MCI along high-symmetry line in the slab BZ for OBCs in the $x$-direction (also with $L=50$) and PBCs in other directions ($(m,t,\triangle)=(+1,-0.5,+1)$) with $m' = 0.1$, for $t'=0$ (black) and $t' = 0.05$ (red).}
  \label{fig:F4}
\end{figure}

\textit{Effects of symmetry-breaking perturbations}---We now consider effects of weak symmetry-breaking terms, which break the tensor product structure---but do not close the bulk energy gap---of the child Bloch Hamiltonian. We first consider a symmetry-breaking term, which is constant in momentum, of the form $\mathcal{H}'_{const} = m' (\tau_z \sigma_z + \tau_z \sigma_0 + \tau_0 \sigma_z + \tau_x \sigma_z + \tau_y \sigma_0)$, with free parameter $m'$. Finite $m'$ reduces the symmetry of the Hamiltonian to class A~\cite{schnyder2008, Ryu_2010}. Rather than gapping out the slab spectrum, this constant term shifts quadratic band-touchings on each edge up and down in energy, respectively, as shown in Fig.~\ref{fig:F4} (a) in black. The quadratic band-touchings are not gapped out, due to differences in location of probability density peaks for the two in-gap states localized on a given edge similar to those observed for the multiplicative Kitaev chain~\cite{pal2023multiplicative}.

We therefore consider an additional momentum-dependent symmetry-breaking term $\mathcal{H}'(\boldsymbol{k}) = 2 t' \cos(k_x) \tau_y \sigma_y$, with free parameter $t'$. This term gaps out the quadratic band-touchings at each edge, but linear band-crossings remain as shown in Fig.~\ref{fig:F4} (a) (dashed red lines). These remaining band-crossings are realised by a pair of non-degenerate bands of states localised on opposite edges.

This bulk-boundary correspondence of the MCI with additional finite $\mathcal{H}'_{const}$ and $\mathcal{H}'(\boldsymbol{k})$ is that of a topological skyrmion phase of matter~\cite{qskhe, winterOEPT}. Indeed, computing the  expectation value texture over the 2D BZ for spin representation $S_i = \tau_i \sigma_0$ (or $\tau_0 \sigma_i$), corresponding to one parent DOF (or the other), yields a skyrmion over the BZ with topological charge of $1$. Combining contributions from individual parents then yields the topological charge of $\mathcal{Q}=2$~\cite{SuppMat}. $\langle \tau_z \sigma_y \rangle$ (or $\langle \tau_y \sigma_z \rangle$) for individual edge states is transferred between edges over the interval in $k_y$ bounded by the in-gap band crossings, as shown in Fig.~\ref{fig:F4} (c), with some deviations due to class A symmetry. This is a signature of topological skyrmion phases~\cite{qskhe}. Finally, we note that the symmetry-breaking terms, $\mathcal{H}'_{const}$ and $\mathcal{H}'(\boldsymbol{k})$, also preserve gaplessness in the mixed MCI, as shown in Fig.~\ref{fig:F4} (d).

\textit{Discussion and Conclusion}---We study multiplicative topological phases (MTPs) of the broader quantum skyrmion Hall effect (QSkHE) framework, focusing on multiplicative Chern insulators (MCIs). We introduce three-dimensional (3D) mixed MCIs and study topological response signatures and robustness of MCIs. Time-reversal symmetric flux insertion reveals a $4 \pi$ Aharonov-Bohm (AB) effect of the MCI understood in terms of a Chern-Simons theory within the QSkHE framework~\cite{patil2024} similarly to the FQHE and related states~\cite{kalmeyer1989}. This indicates the MCI exhibits a pairing mechanism yielding a composite compactified many-body state generalising Jain's notion of a composite particle~\cite{Jain:1989tx}. We also observe evidence of an additional topological invariant associated with compactified topologically non-trivial many body states~\cite{patil2024}, however, suggesting richer physics of the QSkHE. Upon breaking the product structure of the bulk Hamiltonian for the MCI, the system evolves adiabatically into more general topological states of the QSkHE~\cite{cook2023, qskhe, patil2024}. Our work therefore reveals MTPs and the QSkHE as a potential framework for unifying understanding of non-interacting and interacting systems.

\textit{Acknowledgements}---We would like to thank R. Moessner for careful reading of the manuscript and many helpful discussions.

\bibliography{main.bib}

\pagebreak

%%%%%%%%%% Merge with supplemental materials %%%%%%%%%%
%\pagebreak
%\widetext
%%%%%%%%%% Prefix a "S" to all equations, figures, tables and reset the counter %%%%%%%%%%
%\setcounter{equation}{0}
%\setcounter{figure}{0}
%\setcounter{table}{0}
%\setcounter{page}{1}
%\makeatletter
%\renewcommand{\theequation}{S\arabic{equation}}
%\renewcommand{\thefigure}{S\arabic{figure}}
%\renewcommand{\bibnumfmt}[1]{[S#1]}
%\renewcommand{\citenumfont}[1]{S#1}
%%%%%%%%%% Prefix a "S" to all equations, figures, tables and reset the counter %%%%%%%%%%
%\appendix

\clearpage

%%%%%%%%% Prefix a "S" to all equations, figures, tables and reset the counter %%%%%%%%%%
\makeatletter
\renewcommand{\theequation}{S\arabic{equation}}
\renewcommand{\thefigure}{S\arabic{figure}}
\renewcommand{\thesection}{S\arabic{section}}
\setcounter{equation}{0}
\setcounter{section}{0}
\onecolumngrid

%%%%%%%%% Prefix a "S" to all equations, figures, tables and reset the counter %%%%%%%%%%

\begin{center}
  \textbf{\large Supplemental material for ``Multiplicative Chern insulator''}\\[.2cm]
  Archi Banerjee$^{1,2,3}$ and Ashley M. Cook$^{1,2,*}$\\[.1cm]
  {\itshape ${}^1$Max Planck Institute for Chemical Physics of Solids, Nöthnitzer Strasse 40, 01187 Dresden, Germany\\
  ${}^2$Max Planck Institute for the Physics of Complex Systems, Nöthnitzer Strasse 38, 01187 Dresden, Germany\\
  ${}^3$SUPA, School of Physics and Astronomy, University of St Andrews, St Andrews KY16 9SS, United Kingdom\\}
  ${}^*$Electronic address: cooka@pks.mpg.de\\
(Dated: \today)\\[1cm]
\end{center}

\section{Computation of skyrmion number}

Here, we compute the skyrmion number $Q$ of the MCI given eigenstates of parent Hamiltonians $1$ and $2$ for the case realizing the $4 \pi$ Aharonov Bohm effect, to further investigate its origin. We consider two-band parent CI systems, and label the occupied and unoccupied states of system $1$ as $\phi(k)_{p,-}$ and $\phi(k)_{p,+}$. We label the occupied and unoccupied states of system $2$ as $\phi(k)_{h,-}$ and $\phi(k)_{h,+}$. Occupied states of the child Hamiltonian are then
\begin{align}
\psi_0 &= \phi(k)_{p,-}\otimes \phi(k)_{h,+}  = \phi(k)_{p,-} \otimes \phi(k)_{p,-} \\
\psi_1 &= \phi(k)_{p,+}\otimes \phi(k)_{h,-}  = \phi(k)_{h,-} \otimes \phi(k)_{h,-} .
\end{align}

$\psi_0$ and $\psi_1$ are degenerate, yet the symmetry-protected tensor product structure yields a $U(1) \times U(1)$ holonomy.

The spin expectation value to compute the skyrmion number is then effectively computed using spin operators for parent 1, $\sigma_p$, and parent 2, $\sigma_h = -\sigma_p$, in terms of spin expectation values of the parents
\begin{align}
\langle s_{p,i} \rangle &= \langle \phi_{p,-} | \sigma_{p,i}| \phi_{p,-}  \rangle \\
\langle s_{h,i} \rangle &= \langle \phi_{h,-} | \sigma_{h,i} | \phi_{h,-}  \rangle.
\end{align}
The spin curvatures for systems $1$ and $2$ are then
\begin{align}
    \Omega_{1,2}(k) &= \langle s_{p/h}(k) \rangle \cdot \left( \partial_{k_x} \langle s_{p/h}(k) \rangle \times \partial_{k_y} \langle s_{p/h}(k) \rangle\right).
\end{align}

However, spin curvatures of system $1$ and $2$ combine additively based on the definition of the total (pseudo)spin~\cite{pal2023multiplicative, pal_multsemimetal} to yield total spin curvature of
\begin{align}
    \Omega_c(k) &= \Omega_1(k) + \Omega_2(k) = 2 \Omega_1(k).
\end{align}

The spin skyrmion number is then
\begin{align}
   Q= \left(1 /4 \pi\right)\int d^2k \Omega_c(k) = 2C_p = 2.
\end{align}
This invariant for the emergent (pseudo)spin charge gauge field, in combination with effective single-particle invariants $C_{p(h)} = 1 (-1)$, gives $\nu = 1/2$ for the composite particles in agreement with numerics.

\section{Characterisation of Tr[\textbf{C} ]}

\begin{figure}[htb!]
  \centering
   \includegraphics[width=.65\textwidth]{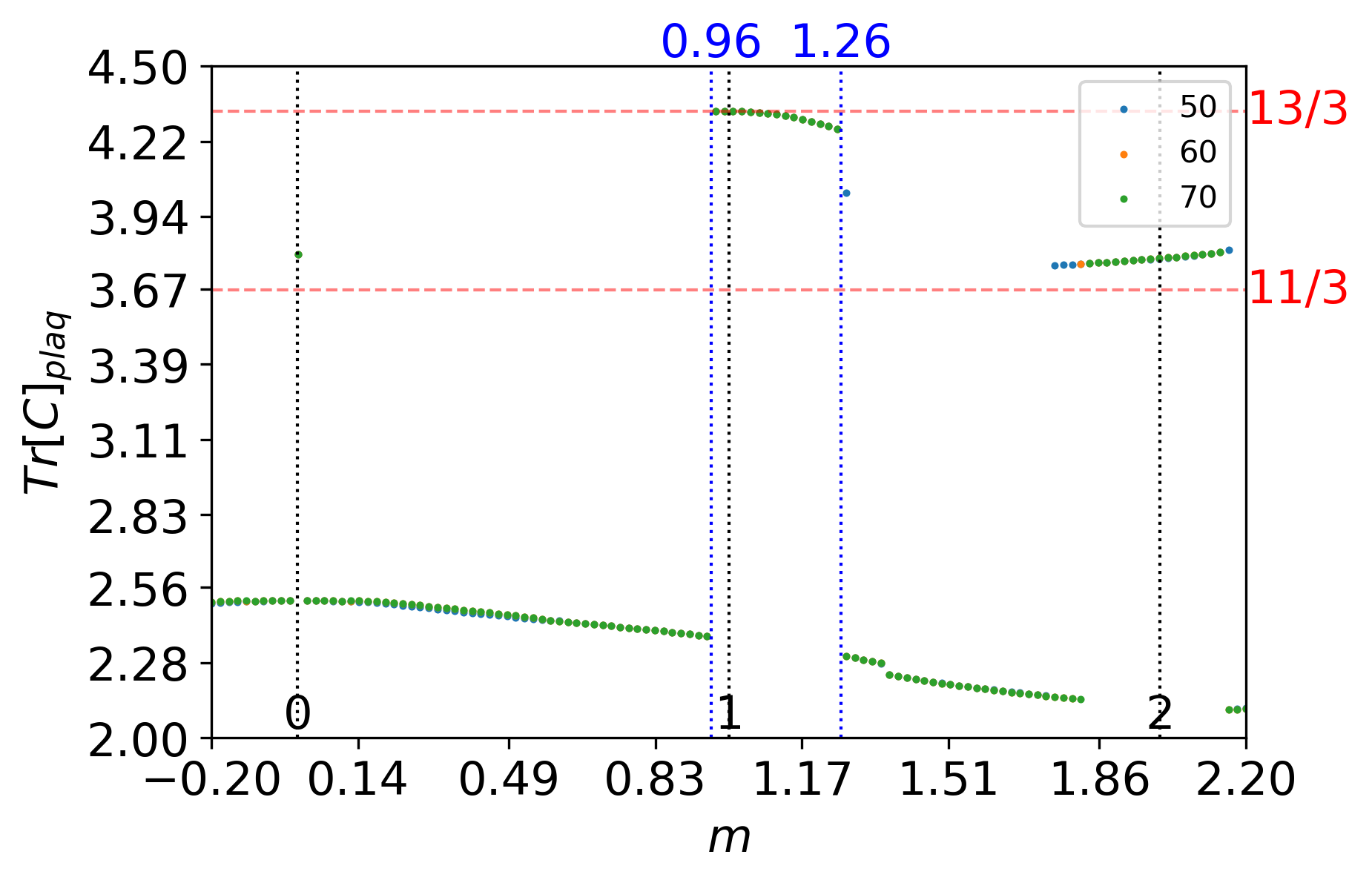}
  \caption{$Tr\left[\boldsymbol{C} \right]$ vs. mass $m$ for system sizes of $50 \times 50$, $60 \times 60$, and $70 \times 70$ unit cells, specifically for a site of magnetic flux insertion, and $\phi=\phi_0$. Orange dashed lines mark values of $13/3$ and $11/3$. Blue dashed lines mark values of $m$ bounding plateau near $13/3$. Black dashed lines mark $m$ values of $0$, $1$, and $2$.}
  \label{supp:charge}
\end{figure}

\begin{figure}[htb!]
  \centering
   \includegraphics[width=.65\textwidth]{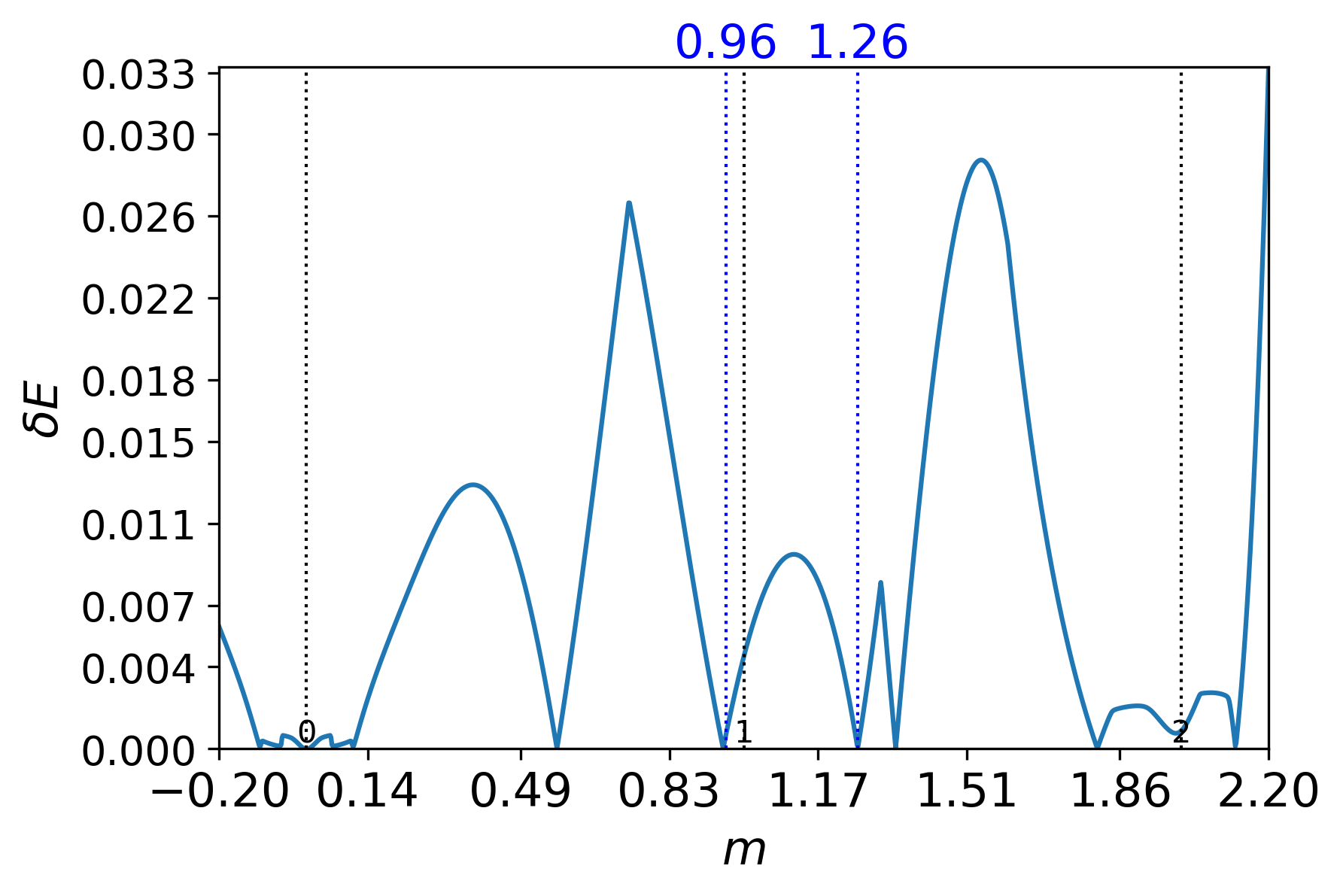}
  \caption{Minimum energy gap $\delta E$ vs. $m$, for $\phi=\phi_0$. Blue dashed lines mark values of $m$ bounding plateau near $13/3$. Black dashed lines mark $m$ values of $0$, $1$, and $2$.}
  \label{supp:gap}
\end{figure}

Here, we provide additional details on how the quantity $Tr\left[\boldsymbol{C} \right]$ varies over phase space. For a location of flux insertion in the lattice $x=20$, $y=25$ and flux $\phi=\phi_0$, we plot $Tr\left[\boldsymbol{C} \right]$ vs. mass $m$ in Fig.~\ref{supp:charge}. A plateau at $13/3$ is observed near $m=1$, with some deviation as $m$ is increased. The bulk energy gap $\delta E$ vs. $m$ for $\phi=\phi_0$ is shown in Fig.~\ref{supp:gap}: the plateau in $Tr\left[\boldsymbol{C} \right]$ is bounded by values of $m$ at which $\delta E = 0$. We also show the determinant of $S_z$ of the occupied subspace, or $S_z^{occ}$ in the main text, vs. $m$, which is expected to stabilise the proposed topological invariant $Tr\left[\boldsymbol{C}\right]$. The plateau near $m=1$ with $13/3$ quantisation corresponds to finite determinant, while other intervals in $m$ have zero determinant.

\begin{figure}[htb!]
  \centering
   \includegraphics[width=.65\textwidth]{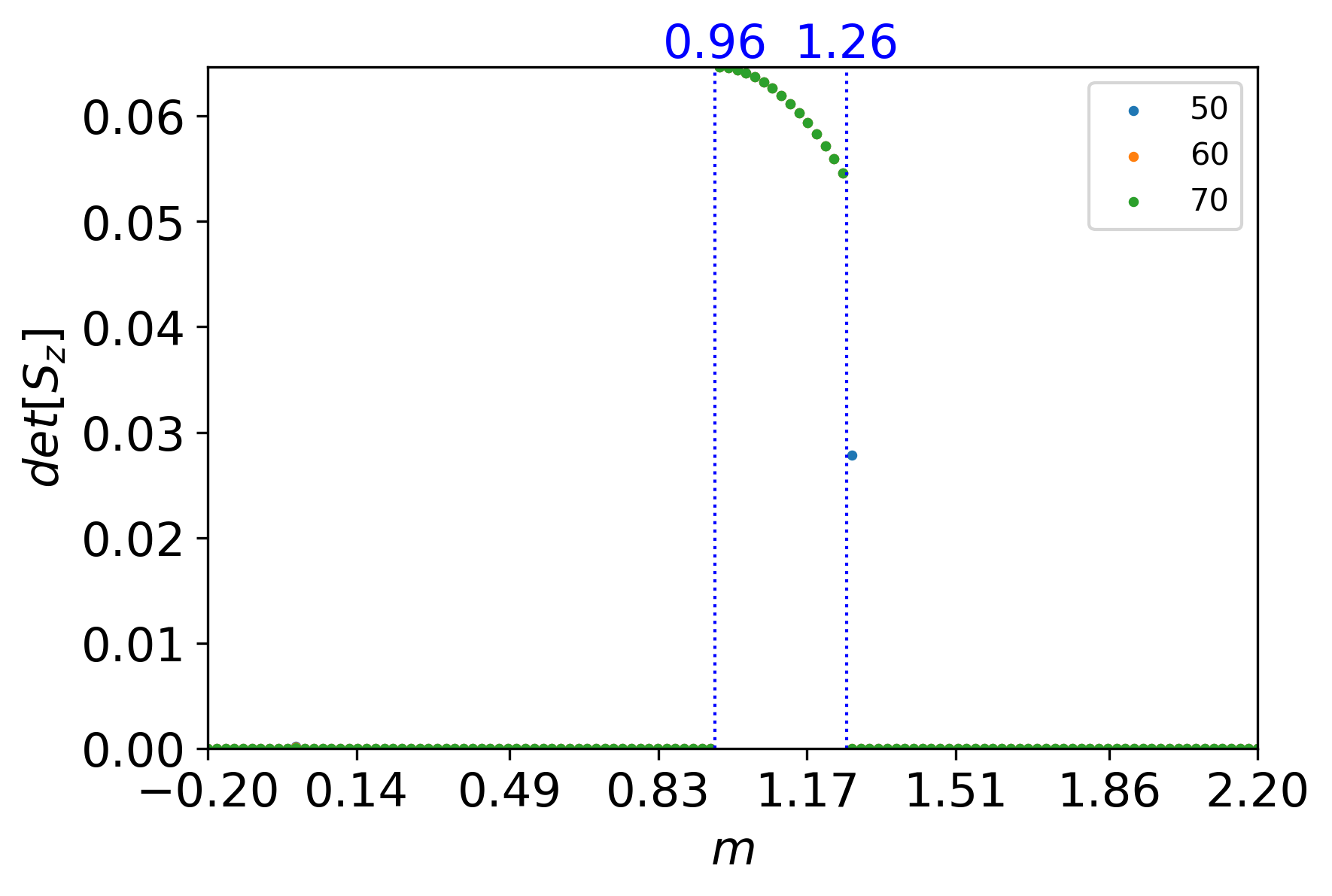}
  \caption{Determinant of the $S_z$ spin operator of the occupied subspace, $S^{occ}_z$, vs. $m$, for $\phi=\phi_0$. Blue dashed lines mark values of $m$ bounding plateau near $13/3$.}
  \label{supp:spinZ}
\end{figure}

%\bibliography{main.bib}

\end{document}